\documentclass[aps,prl,times,twocolumn,superscriptaddress,showpacs,floatfix]{revtex4-1}

\usepackage{mathptmx}

\usepackage[T1]{fontenc}
\usepackage[latin9]{inputenc}
\usepackage{amsmath}
\usepackage{graphicx}
\usepackage[multiple]{footmisc}

\begin{document}

\title{Nature of Partial Magnetic Order in the Frustrated Antiferromagnet Gd$_{2}$Ti$_{2}$O$_{7}$}

\author{Joseph A. M. Paddison}
\email{paddisonja@ornl.gov}
\affiliation{Materials Science and Technology Division, Oak Ridge National Laboratory, Oak Ridge, TN 37831, USA}
\affiliation{Churchill College, University of Cambridge, Storey's Way, Cambridge CB3 0DS, U.K.}
\affiliation{Department of Chemistry, University of Oxford, Inorganic Chemistry Laboratory, South Parks Road, Oxford OX1 3QR, U.K.}
\affiliation{ISIS Neutron and Muon Source, Rutherford Appleton Laboratory, Didcot OX11 0QX, U.K.}

\author{Georg Ehlers}
\affiliation{Neutron Technologies Division, Oak Ridge National Laboratory, Oak Ridge, TN 37831, USA}

\author{Andrew B. Cairns}
\affiliation{Department of Chemistry, University of Oxford, Inorganic Chemistry Laboratory, South Parks Road, Oxford OX1 3QR, U.K.}

\author{Jason S. Gardner}
\affiliation{Songshan Lake Materials Laboratory, Dongguan, Guangdong 523808, China}

\author{Oleg A. Petrenko}
\affiliation{Department of Physics, University of Warwick, Coventry CV4 7AL, U.K.}

\author{Nicholas P. Butch}
\affiliation{NIST Center for Neutron Research, National Institute of Standards and Technology, Gaithersburg, MD 20899, USA}

\author{Dmitry D. Khalyavin}
\affiliation{ISIS Neutron and Muon Source, Rutherford Appleton Laboratory, Didcot OX11 0QX, U.K.}

\author{Pascal Manuel}
\affiliation{ISIS Neutron and Muon Source, Rutherford Appleton Laboratory, Didcot OX11 0QX, U.K.}

\author{Henry E. Fischer}
\affiliation{Institut Laue-Langevin, 71 Avenue des Martyrs, CS 20156, 38042 Grenoble Cedex 9, France}

\author{Haidong Zhou}
\affiliation{Department of Physics and Astronomy, University of Tennessee, Knoxville, TN 37996, USA}
\affiliation{National High Magnetic Field Laboratory, Florida State University, Tallahassee, FL 32310, USA}

\author{Andrew L. Goodwin}
\affiliation{Department of Chemistry, University of Oxford, Inorganic Chemistry Laboratory, South Parks Road, Oxford OX1 3QR, U.K.}

\author{J. Ross Stewart}
\email{ross.stewart@stfc.ac.uk}
\affiliation{ISIS Neutron and Muon Source, Rutherford Appleton Laboratory, Didcot OX11 0QX, U.K.}

\date{\today}

\begin{abstract}
Partially-ordered magnets are distinct from both spin liquids and conventional ordered magnets because order and disorder coexist in the same magnetic phase. Here, we determine the nature of partial order in the canonical frustrated pyrochlore antiferromagnet Gd$_{2}$Ti$_{2}$O$_{7}$. Using single-crystal neutron-diffraction measurements in applied magnetic field, magnetic symmetry analysis, inelastic neutron-scattering measurements, and spin-wave modeling, we show that its low-temperature magnetic structure involves two propagation vectors (2-$\mathbf{k}$ structure) with suppressed ordered magnetic moments and enhanced spin-wave fluctuations. Our experimental results support theoretical predictions of thermal fluctuation-driven order in Gd$_{2}$Ti$_{2}$O$_{7}$.

\end{abstract}

\maketitle

Geometrical frustration is a central theme of condensed-matter physics because it can generate exotic magnetic states. These states can typically be divided into spin liquids, in which frustration inhibits long-range magnetic order, and spin solids, in which perturbations to the dominant frustrated interactions drive magnetic order \cite{Moessner_2006}. Defying this classification, some frustrated magnets exhibit \emph{partial} magnetic order \cite{Movshovich_1999,Greedan_2002,Zheng_2005,Rule_2007,Cao_2007,Ehlers_2007,Hayes_2011}---the coexistence of order and disorder in the same magnetic phase. Magnetic partial order can be driven by fluctuations in an ``order-by-disorder'' scenario \cite{Javanparast_2015}, by interactions between emergent degrees of freedom in spin-fragmented states \cite{Brooks-Bartlett_2014,Paddison_2016,Petit_2016}, or by proximity to a quantum critical point \cite{Pfleiderer_2004}, while structural partial order can drive the behavior of materials such as fast-ion conductors \cite{Rice_1974,Keen_1996}, Pb-based photovoltaics \cite{Weller_2015,Eames_2015}, and high-pressure elemental phases \cite{Gregoryanz_2008}. To benchmark theories of partially-ordered states \cite{Chern_2008,Javanparast_2015,Wills_2006}, experimental determination of the nature of partial order in real materials is crucial.

\begin{figure}[hhh!]
\begin{center}
\includegraphics[scale=1]{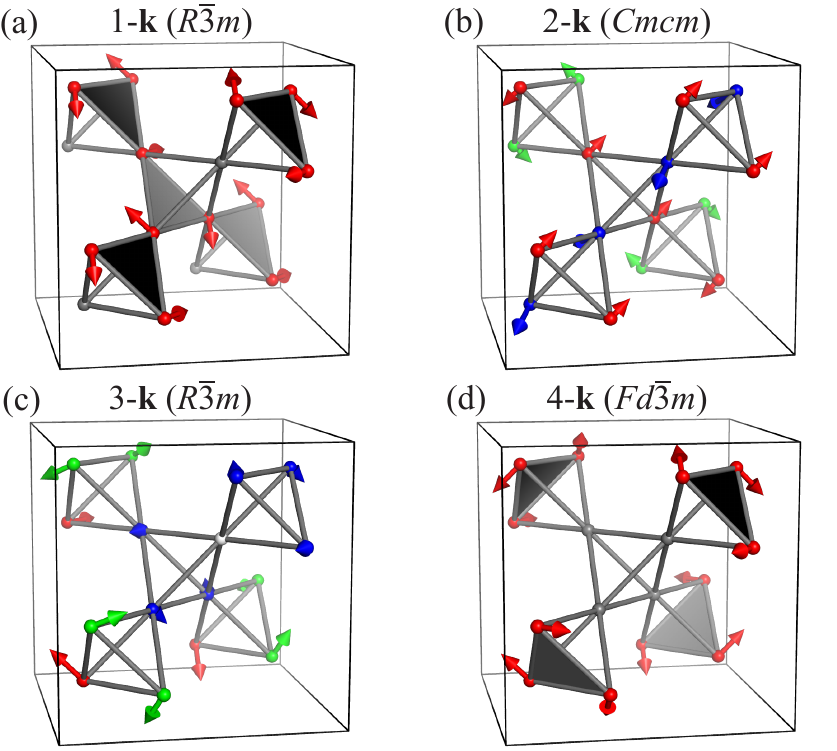}
\end{center}
\caption{\label{fig1} Basic candidate structures for Gd$_{2}$Ti$_{2}$O$_{7}$, showing (a) 1-$\mathbf{k}$, (b) 2-$\mathbf{k}$, (c) 3-$\mathbf{k}$, and (d) 4-$\mathbf{k}$ structures. The symmetry of each structure is labeled. A single crystallographic unit cell is shown in each case; spin orientations are reversed in adjacent unit cells because of the $\mathbf{k}=(\frac{1}{2} \frac{1}{2} \frac{1}{2})$ propagation vector. Spin directions are shown as arrows with lengths proportional to the ordered moment magnitude $\mu_\mathrm{ord}$. Arrows of different colors indicate symmetry-inequivalent magnetic sites, and paramagnetic sites with zero $\mu_\mathrm{ord}$ in (a) and (d) are shown as grey spheres.}\end{figure}

Materials in which magnetic ions occupy a pyrochlore lattice of corner-sharing tetrahedra provide opportunities for realizing exotic frustrated states \cite{Gardner_2010}. The frustrated pyrochlore antiferromagnet Gd$_{2}$Ti$_{2}$O$_{7}$ is a canonical partially-ordered system in which magnetic Gd$^{3+}$ ions ($S=7/2$) undergo two phase transitions at $T_{1} = 1.1$\,K and $T_{2} = 0.75$\,K \cite{Raju_1999,Ramirez_2002,Bonville_2003,Petrenko_2004,Petrenko_2011}. Both low-temperature (LT; $T\ll T_2$) and intermediate ($T_2 <T<T_1$) phases have magnetic propagation vector $\mathbf{k}=(\frac{1}{2} \frac{1}{2} \frac{1}{2})$ \cite{Champion_2001,Stewart_2004} and are partially ordered, as shown by the coexistence of magnetic Bragg and diffuse scattering in polarized-neutron scattering measurements \cite{Stewart_2004}. However, the LT magnetic structure of Gd$_{2}$Ti$_{2}$O$_{7}$ has not yet been conclusively solved, for two reasons. First, the large neutron absorption cross-section of natural Gd makes neutron-scattering experiments on large crystals challenging. Second, most experimental probes are unable to distinguish a magnetic structure that orders with a single $\mathbf{k}=(\frac{1}{2} \frac{1}{2} \frac{1}{2})$ wavevector (``1-$\mathbf{k}$ structure'') from structures that superpose symmetry-equivalent $\mathbf{k}\in \langle \frac{1}{2} \frac{1}{2} \frac{1}{2} \rangle$ ($n$-$\mathbf{k}$ structures, where $n\in \{2,3,4\}$ in this case). Four such candidate structures for Gd$_{2}$Ti$_{2}$O$_{7}$ are shown in Fig.~\ref{fig1}. All are partially ordered, but each has a different modulation of the ordered magnetic moment $\mu_{\mathrm{ord}}$: the 1-$\mathbf{k}$ and 4-$\mathbf{k}$ structures have 25\% interstitial paramagnetic sites, whereas 2-$\mathbf{k}$ and 3-$\mathbf{k}$ have more complicated $\mu_{\mathrm{ord}}$ modulations. It was proposed in Ref.~\onlinecite{Stewart_2004} that magnetic diffuse-scattering measurements support a 4-$\mathbf{k}$ structure with cubic magnetic symmetry. However, this result was called into question by the observation of transverse magnetization in small applied magnetic fields $\mathbf{H} \parallel \langle 112 \rangle$ and $\langle 100 \rangle$, which is inconsistent with cubic symmetry \cite{Glazkov_2007,Petrenko_2012}. Moreover, theoretical modeling suggests that, while the intermediate structure is 4-$\mathbf{k}$ and stabilized by fluctuations, the LT ground state may actually be 2-$\mathbf{k}$ \cite{Javanparast_2015}---a striking prediction that has awaited a conclusive experimental test.

In this Letter, we experimentally determine the nature of partial magnetic order in Gd$_{2}$Ti$_{2}$O$_{7}$ using neutron-scattering measurements of isotopically-enriched powder and single-crystal samples, combined with symmetry analysis and spin-wave calculations. We show that the LT state of Gd$_{2}$Ti$_{2}$O$_{7}$ is actually 2-$\mathbf{k}$, in agreement with theory \cite{Javanparast_2015} but in contradiction with the interpretation of previous experiments \cite{Stewart_2004}. Our paper is structured as follows. We first present single-crystal neutron-diffraction measurements in applied magnetic field that indicate non-cubic magnetic symmetry. We then perform a comprehensive symmetry analysis of candidate magnetic structures. Finally, we show that only a 2-$\mathbf{k}$ structure is consistent with LT inelastic neutron-scattering (INS) data. 

Measurements of magnetic Bragg intensities in zero applied magnetic field do not directly distinguish the structures shown in Fig.~\ref{fig1}, due to spherical averaging in powder samples and averaging over degenerate magnetic domains in single crystals---a phenomenon known as the ``multi-$\mathbf{k}$ problem'' \cite{Kouvel_1963}. To address this problem, we performed neutron-diffraction measurements on a $\sim$10\,$\mathrm{mm^{3}}$ single crystal using the WISH diffractometer at ISIS \cite{Chapon_2011}, and applied a weak magnetic field $\mathbf{H}\parallel [1\bar{1}0]$ to break the domain degeneracy at $T=0.07$\,K after zero-field cooling. The sample was cut from a larger crystal prepared by the floating-zone image furnace method \cite{Balakrishnan_1998,Gardner_1998} and was 99.4\% enriched with $^{160}$Gd to minimize absorption. Domains of the cubic 4-$\mathbf{k}$ structure are related only by translational and time-reversal symmetries and hence appear identical to neutrons, whereas domains of other $n$-$\mathbf{k}$ structures have different diffraction patterns. A field-induced domain imbalance is therefore expected to leave the diffraction pattern unchanged \emph{only} if the LT structure is 4-$\mathbf{k}$.

\begin{figure}
\begin{center}
\includegraphics[scale=1]{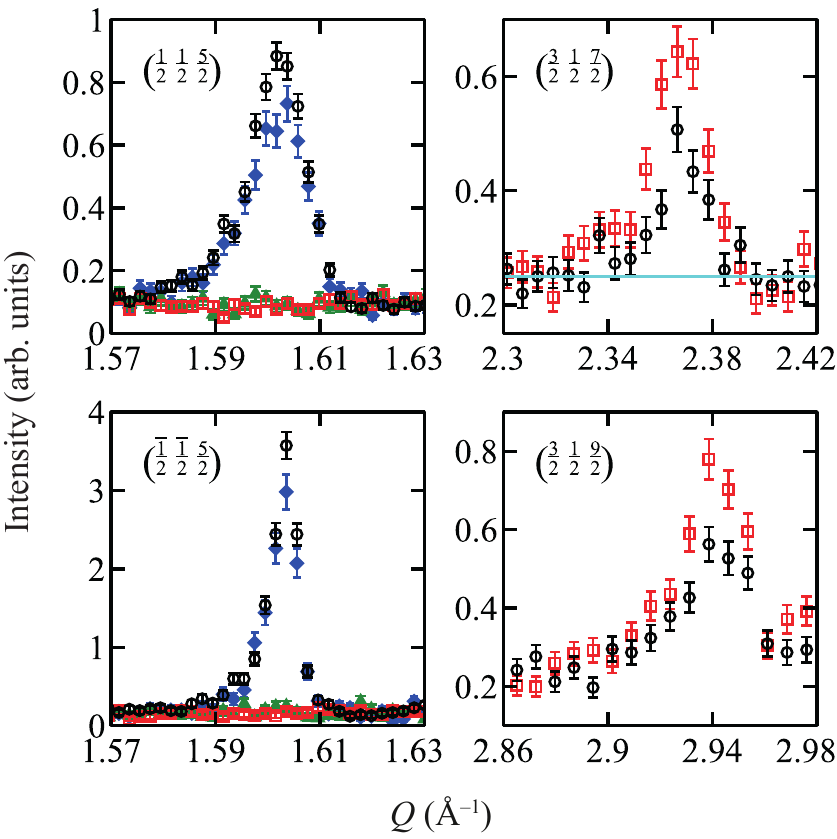}
\end{center}
\caption{\label{fig2} Intensities of selected single-crystal magnetic Bragg peaks at different values of applied magnetic field $\mathbf{H}\parallel [1\bar{1}0]$ at $T = 0.07$\,K. Magnetic Bragg peaks are labelled in each panel; left-hand panels show peaks within the $(hhl)$ plane and right-hand panels show peaks outside the $(hhl)$ plane. Points are colored as follows: $\mu_{0}\mathbf{H}=0$ (black hollow circles), $\mu_{0}\mathbf{H}=0.1$\,T (blue filled diamonds), $\mu_{0}\mathbf{H}=0.2$\,T (green filled triangles), and $\mu_{0}\mathbf{H}=0.5$\,T (red hollow squares). For clarity, only $\mu_{0}\mathbf{H}=0$ and $\mu_{0}\mathbf{H}=0.5$\,T are shown in the right-hand panels. For the $(\frac{3}{2} \frac{3}{2} \frac{7}{2})$ peak, the integrated intensity is 0.11(2) units for $H=0$ and 0.22(2) units for $\mu_{0}\mathbf{H}=0.5$\,T, consistent with the expected doubling (the estimated background is shown as a light blue line). 
}\end{figure}

The magnetic field dependence of selected magnetic Bragg intensities is shown in Fig.~\ref{fig2}. Magnetic Bragg peaks in the $(hhl)$ plane disappear in small applied field $0.2\,\leq \mu_{0}H\leq 0.5$\,T, while the intensity of magnetic Bragg peaks outside the $(hhl)$ plane increases. These observations are incompatible with the cubic 4-$\mathbf{k}$ structure, unless the applied field actually caused a magnetic phase
transition rather than a domain imbalance. This scenario occurs in Er$_2$Ti$_2$O$_7$ \cite{Ruff_2008}, but is unlikely in Gd$_2$Ti$_2$O$_7$, in which there is no
experimental evidence for such a phase transition in either specific heat \cite{Petrenko_2004} or torque magnetometry \cite{Glazkov_2007} measurements
for $\mathbf{H}\parallel \langle110\rangle$ of less than $2\,\mathrm{T}$ at base temperature. 
The field-induced uniform magnetization is also too small to suppress magnetic Bragg peaks fully for $0.2\,\leq \mu_{0}H\leq 0.5$\,T ($M\approx 0.2\,\mu_\mathrm{B}$ for $\mu_{0}H\approx 0.2$\,T \cite{Petrenko_2011,Petrenko_2012}). These observations strongly disfavor the cubic 4-$\mathbf{k}$ structure, but do not distinguish candidate non-cubic structures.
 
\begin{figure}
\begin{center}
\includegraphics[scale=1]{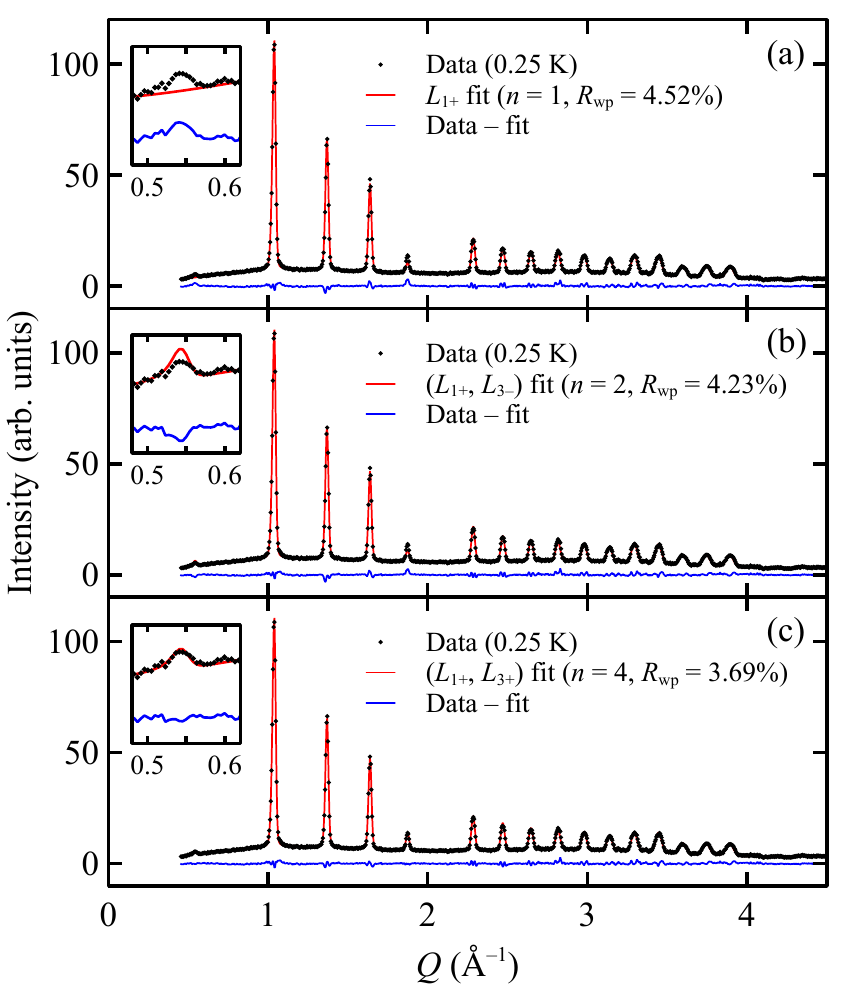}
\end{center}
\caption{\label{fig3} Experimental powder neutron-diffraction data (from Ref.~\onlinecite{Stewart_2004}, $\lambda=2.42$\,\AA) in the LT phase (0.25\,K) and Rietveld fits for (a) the $\mathrm{L_{1+}}$ irrep, (b) the $\left(\mathrm{L_{1+}},\mathrm{L_{3-}}\right)$ irrep pair, and (c) and the $\left(\mathrm{L_{1+}},\mathrm{L_{3+}}\right)$ irrep pair. Experimental data are shown as black points, fits as red lines, and data--fit as blue lines. For each model, the number of free parameters (magnetic distortion modes) $n$ and the goodness-of-fit metric $R_\mathrm{wp}$ are shown. The insets show the $(\frac{1}{2} \frac{1}{2} \frac{1}{2})$ magnetic Bragg peak on an expanded scale. }\end{figure}

To constrain further the LT magnetic structure, we reinterpret published powder-neutron diffraction data \cite{Stewart_2004} using a comprehensive symmetry analysis. Fig.~\ref{fig3} shows data collected in the LT phase (0.25\,K) using the D20 diffractometer at the ILL \cite{Hansen_2008}. Nearly all the features of the data can be modeled using a single magnetic irreducible representation (``irrep''), denoted $L_{1+}$ in Miller and Love's notation [Fig.~\ref{fig3}(a)] \cite{Cracknell_1979}. The $L_{1+}$ model is the best currently available for Gd$_2$Ti$_2$O$_7$, and generates the four basic $n$-$\mathbf{k}$ structures shown in Fig.~\ref{fig1}. Crucially, however, the $(\frac{1}{2} \frac{1}{2}  \frac{1}{2} )$ magnetic Bragg peak observed in the 0.25\,K experimental data is absent for the $L_{1+}$ model [inset to Fig.~\ref{fig3}(a)]. This result suggests that, while the $L_{1+}$ irrep is the main contributor to the LT magnetic structure, at least one other irrep must also be present. We therefore tested the four possible combinations of $L_{1+}$ with one other irrep ($L_{2+}$, $L_{3+}$, $L_{1-}$, or $L_{3-}$). Only the $(L_{1+},L_{3-})$ and $(L_{1+},L_{3+})$ combinations allow nonzero intensity of the $(\frac{1}{2} \frac{1}{2}  \frac{1}{2} )$ peak and so are candidates. For each of these irrep pairs, we generated structural models using the Isodistort program \cite{Campbell_2006,Stokes_2006}, and treated the magnetic distortion-mode amplitudes as free parameters which we optimized against our diffraction data using Topas Academic \cite{Coelho_2012}.

Fig.~\ref{fig3} compares fits to diffraction data for the single-irrep $L_{1+}$ model with the two-irrep $(L_{1+},L_{3-})$ and $(L_{1+},L_{3+})$ models. The two-irrep models yield an improved overall fit to the data---in particular, to the $(\frac{1}{2} \frac{1}{2}  \frac{1}{2} )$ peak---by increasing the number of free parameters as indicated in Fig.~\ref{fig3}. 
Unfortunately, the inclusion of an additional irrep also increases the number of candidate structures from four to 32 (see SM), all consistent with powder-diffraction data. We therefore apply the criterion that no site should have $\mu_{\mathrm{ord}} > 7.0\,\mu_{\mathrm{B}} $---the maximum value for $S=7/2$ Gd$^{3+}$ ions---which reduces the number of candidate structures to eight. Four of these are monoclinic $(L_{1+},L_{3-})$ variants of the 1-$\mathbf{k}$ structure, in which paramagnetic spins order with a small $\mu_{\mathrm{ord}}$ \cite{Stewart_2004}. However, these structures are disfavored by symmetry because the $L_{3-}$ irrep is not a symmetry-allowed secondary order parameter (SOP) here \cite{Campbell_2006,Stokes_2006}. The other four comprise three 2-$\mathbf{k}$ structures and one 4-$\mathbf{k}$ structure, of which only two $(L_{1+},L_{3+})$ $2$-$\mathbf{k}$ structures contain a symmetry-allowed SOP. These two-irrep structures are consistent with the moment-length constraint, but the single-irrep 2-$\mathbf{k}$ structure is not, and was therefore neglected in previous work \cite{Stewart_2004}.
Importantly, however, the moment-length constraint rules out all 3-$\mathbf{k}$ structures, so we do not consider these further. 

\begin{figure*}
\begin{center}
\includegraphics[scale=1]{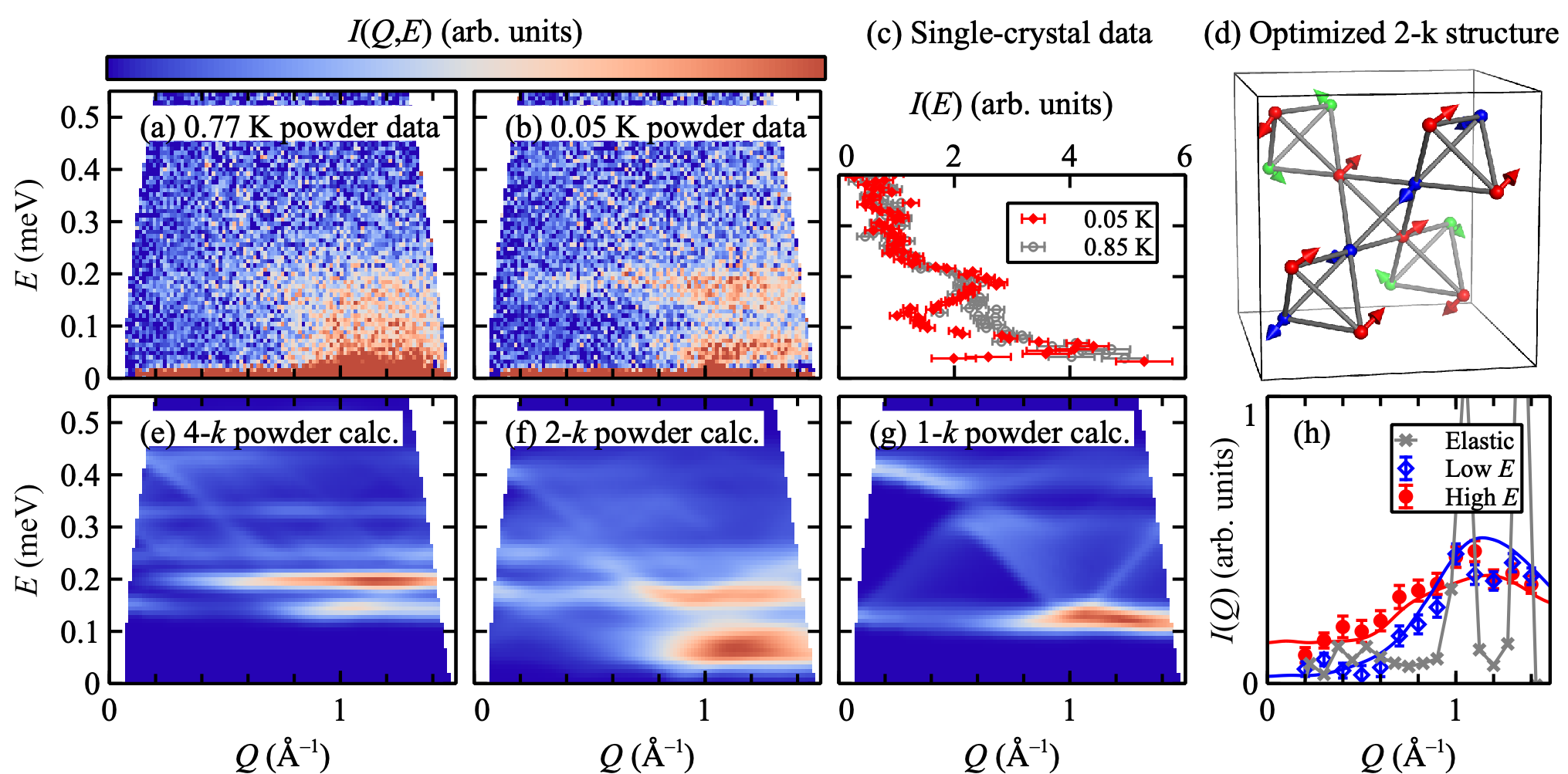}
\end{center}
\caption{\label{fig4} (a) Inelastic neutron scattering data collected on a powder sample in the intermediate phase (0.77\,K). (b) Powder INS data collected in the LT phase ($\sim$0.05\,K). (c) Single-crystal INS data integrated over $(h,k,l)=(0\pm1,0\pm1,\frac{3}{2}\pm\frac{1}{2})$ in the LT phase (0.05\,K, solid red diamonds) and intermediate phase (0.85\,K, empty grey circles). (d) Optimized 2-$\mathbf{k}$ magnetic structure. (e) Powder-averaged linear spin-wave theory (LSWT) calculation for the $L_{1+}$ 4-$\mathbf{k}$ structure. (f) Powder LSWT calculation for the optimized $(L_{1+},L_{3+})$ 2-$\mathbf{k}$ structure discussed in the text. (g) Powder LSWT calculation for the $L_{1+}$ 1-$\mathbf{k}$ structure. (h) Powder INS data and spin-wave calculations at 0.05\,K. Grey crosses show the measured elastic line ($-0.03<E<0.03$\,meV; vertically shifted for clarity), empty blue diamonds show the measured low-$E$ mode ($0.03\leq E<0.12$\,meV), and filled red circles show the measured high-$E$ mode ($0.12\leq E\leq 0.21$\,meV). Spin-wave calculations for low-$E$ and high-$E$ modes are shown as blue and red lines, respectively, and have been vertically scaled by the same factor to match the experimental data.}\end{figure*}

Since magnetic Bragg scattering does not fully distinguish candidate structures, we turn to high-resolution INS experiments. Measurements were performed using the DCS spectrometer at NIST \cite{Copley_2003} on a $\sim$0.2\,g portion of the powder sample studied in Ref.~\onlinecite{Stewart_2004}. An incident wavelength of $8$\,\AA~yielded an energy resolution $\approx0.025$\,meV (FWHM). Figs.~\ref{fig4}(a) and \ref{fig4}(b) show background-subtracted powder INS data in the intermediate phase (0.77\,K) and the LT phase ($\sim$0.05\,K), respectively. The magnetic scattering at 0.77\,K is broad in $Q$ and $E$. By contrast, the LT data show two relatively flat modes at energies of approximately 0.06\,meV and 0.17\,meV. We performed additional INS measurements on a thin piece of our single crystal using the CNCS spectrometer at ORNL, which show that the background-subtracted single-crystal scattering integrated over $(h,k,l)=(0\pm1,0\pm1,\frac{3}{2}\pm\frac{1}{2})$ resembles the powder data [Fig.~\ref{fig4}(c)]. Our observation of a mode at very low energy (0.06\,meV) is consistent with observations of low-energy dynamics in neutron spin echo \cite{Ehlers_2006} and muon-spin rotation  \cite{Yaouanc_2005,Dunsiger_2006} experiments.

We use linear spin-wave theory to test models against the LT excitation spectrum. The minimal spin Hamiltonian for Gd$_2$Ti$_2$O$_7$ is given by

\begin{equation}
H={J_1}\sum_{\langle i,j \rangle}{\mathbf{S}}_{i}\cdot{\mathbf{S}}_{j}+{J_2}\sum_{\langle\langle i,j \rangle\rangle}{\mathbf{S}}_{i}\cdot{\mathbf{S}}_{j}+D\sum_{i}({S}^{z}_{i})^2+H_\mathrm{dip},
\label{eq:hamiltonian}
\end{equation}
where $J_1$ and $J_2$ denote Heisenberg exchange interactions between nearest neighbor and next-nearest neighbor spin pairs, which are denoted by angle brackets $\langle \,\rangle$ and $\langle \langle\, \rangle \rangle$, respectively; $D$ is a single-ion anisotropy term that arises from mixture of the excited $^{6}S_{7/2}$ atomic state into the $^{8}S_{7/2}$ ground state \cite{Glazkov_2005}; and $H_\mathrm{dip}$ is the long-ranged magnetic dipolar interaction with energy scale $D_{\mathrm{dip}}S(S+1)=0.84$\,K \cite{Raju_1999}.
We take $J_{1}S(S+1)=4.8$\,K \cite{Raju_1999}, and include a small ferromagnetic $J_{2}=-0.04J_{1}$ to stabilize  $\mathbf{k}=(\frac{1}{2} \frac{1}{2} \frac{1}{2})$ ordering \cite{Raju_1999,Wills_2006,Javanparast_2015}. Electron-spin resonance (ESR) experiments find an easy-plane anisotropy that favors spin alignment perpendicular to local $\langle 111 \rangle$ axes \cite{Glazkov_2005}; we take $DS^2=1.5$\,K to match our INS data. Further evidence for this Hamiltonian comes from measurements of the paramagnetic ($T>T_1$) diffuse scattering using the D4 diffractometer at the ILL \cite{Fischer_2002}, which are consistent with Monte Carlo simulations \cite{Paddison_2017} for these parameters (see SM).

A prerequisite for spin-wave modeling is that the magnetic structure is a local energy minimum of Eq.~\eqref{eq:hamiltonian}. Using the SpinW program \cite{Toth_2015}, we tested which of the eight candidate structures are proximate to energy minima by iteratively aligning each spin with its mean field and checking for stability \emph{via} the absence of imaginary spin-wave modes. Two candidate structures---one 2-$\mathbf{k}$ and one 4-$\mathbf{k}$---are locally stable; both derive from the $(L_{1+},L_{3+})$ irrep pair that yields the best fit to diffraction data [Fig.~\ref{fig3}(c)]. By contrast, all candidate 1-$\mathbf{k}$ structures with nonzero interstitial $\mu_\mathrm{ord}$ are unstable. Fig.~\ref{fig4}(d) shows the optimized $(L_{1+},L_{3+})$ 2-$\mathbf{k}$ structure, which resembles the refined structure (see SM). Compared to the $L_{1+}$ 2-$\mathbf{k}$ structure shown in Fig.~\ref{fig1}(b), it has canted magnetic moments with more uniform magnitudes, 
$\mu_{\mathrm{ord}}/\mu_{\mathrm{B}} \in \{ 6.1(1),4.6(1),6.2(1) \}$ in a 1:1:2 ratio. However, the suppression of $\mu_{\mathrm{ord}}$ compared to its theoretical value of 7.0\,$\mu_{\mathrm{B}}$ indicates partial ordering.
Figs.~\ref{fig4}(e), \ref{fig4}(f), and \ref{fig4}(g) show the calculated spin-wave spectra for the $L_{1+}$ 4-$\mathbf{k}$, $(L_{1+},L_{3+})$ 2-$\mathbf{k}$, and $L_{1+}$ 1-$\mathbf{k}$ structures, respectively. The 1-$\mathbf{k}$ and 4-$\mathbf{k}$ calculations strongly disagree with the 0.05\,K experimental data. By contrast, the 2-$\mathbf{k}$ calculation reproduces well the experimental data, most importantly the prominent low-energy ($\sim$0.06\,meV) mode. The $Q$ dependences of the low-$E$ ($0.03\leq E<0.12$\,meV) and high-$E$ ($0.12\leq E\leq0.21$\,meV) modes confirm this agreement [Fig.~\ref{fig4}(h)].
We therefore conclude that the $(L_{1+},L_{3+})$ 2-$\mathbf{k}$ structure is the correct LT model.

Our experimental result that the LT structure of Gd$_2$Ti$_2$O$_7$ is 2-$\mathbf{k}$ confirms state-of-the-art theoretical predictions \cite{Javanparast_2015} and solves a longstanding problem in the field of frustrated pyrochlore oxides. However, it contradicts a previous experimental study \cite{Stewart_2004}, which proposed a 4-$\mathbf{k}$ LT structure based on analysis of LT magnetic diffuse scattering. This study did not consider 2-$\mathbf{k}$ structures, because the single-irrep 2-$\mathbf{k}$ structure is unphysical and two-irrep 2-$\mathbf{k}$ structures were not identified. 
It also assumed that ordered sites contribute no diffuse scattering; however, this assumption is incorrect because spin-wave scattering from ordered sites contributes to the energy-integrated diffuse intensity. In a fully-ordered system, $\mu_{\mathrm{ord}}/\mu_{\mathrm{B}}=2S$, so that spin-wave scattering comprises $1/(S+1)=22\%$ of the total intensity for $S=7/2$. By contrast, partially-ordered Gd$_2$Ti$_2$O$_7$ has $\mu_{\mathrm{ord}}/\mu_{\mathrm{B}}<2S$; by combining the refined values of $\mu_{\mathrm{ord}}$ with the total-moment sum rule, we estimate that LT diffuse scattering comprises $46(2)\%$ of the total intensity. Since the elastic scattering ($|E|<0.03$\,meV) is essentially flat away from Bragg peaks, whereas the inelastic scattering shows clear $Q$ dependence [Fig.~\ref{fig4}(h)], the LT diffuse scattering arises primarily from spin-wave fluctuations and not from static spin disorder or relaxational (paramagnetic) spin dynamics. The LT partial ordering of Gd$_2$Ti$_2$O$_7$ is therefore 2-$\mathbf{k}$ with reduced ordered moments and enhanced spin-wave fluctuations. The orthorhombic symmetry of this structure is expected to drive a crystallographic distortion \emph{via} spin-lattice coupling; however, our additional high-resolution powder neutron diffraction measurements do not show clear evidence of peak splitting, and while a statistically-significant rhombohedral distortion could be refined (see SM), orthorhombic refinements were inconclusive due to their increased number of parameters. 
The intermediate-temperature partial ordering is also unusual. A 4-$\mathbf{k}$ structure is predicted to be stabilized by thermal fluctuations at $T_1$ \cite{Javanparast_2015}, consistent with the absence of the $(\frac{1}{2} \frac{1}{2} \frac{1}{2})$ magnetic Bragg peak in the intermediate phase \cite{Stewart_2004}. However, our 0.77\,K INS data show only broad inelastic features, inconsistent with spin-wave predictions. This unexpected result may be explained if periodically-arranged paramagnetic sites in a 4-$\mathbf{k}$ structure suppress propagating spin-wave excitations, similar to the effect of paramagnetic impurities \cite{Brenig_2013}. This intriguing possibility requires further theoretical investigation.


We are grateful to A.~T.~Boothroyd, S.~T.~Bramwell, M.~J.~Cliffe, M.~J.~P.~Gingras, P. McClarty, M.~Mourigal, P.~J.~Saines, and A.~S.~Wills for useful discussions, to O.~Kirichek and the ISIS Sample Environment Group for cryogenic support, and to J.~Makepeace and M.~S. Senn for assistance with \textsc{Topas} 5. J.A.M.P.'s work at ORNL was supported by the Laboratory Directed Research and Development Program of Oak Ridge National Laboratory, managed by UT-Battelle, LLC for the US Department of Energy (manuscript preparation). J.A.M.P.'s work at Cambridge (magnetic structure analysis) was supported by Churchill College, University of Cambridge. J.A.M.P., A.B.C., and A.L.G. acknowledge financial support from the STFC, EPSRC (EP/G004528/2) and ERC (Ref: 279705). A portion of this research used resources at the Spallation Neutron Source, a DOE Office of Science User Facility operated by the Oak Ridge National Laboratory. Work at NHMFL (H.D.Z.) was supported by the NSF-DMR-1157490 and the State of Florida and U.S. Department of Energy. Experiments at the ISIS Neutron and Muon Source were supported by a beam time allocation from the STFC (U.K.).


\end{document}